\documentclass[footinbib, aps,pra,reprint, 10pt,twocolumn,showpacs, superscriptaddress]{revtex4-1}
\usepackage{amsfonts,amsmath,amssymb}
\usepackage{blindtext}
\usepackage{graphicx}
\usepackage[pdfstartview=FitH,colorlinks=true,linkcolor=blue,citecolor=blue,urlcolor=blue]{hyperref}
\usepackage{tikz}
\usepackage{verbatim}

\usepackage{times}

\usepackage[english]{babel}
\usepackage{latexsym}
\usepackage{graphics}
\usepackage{subfigure}
\usepackage{epsfig}
\usepackage{color}
\usepackage{braket}
\usepackage[T1]{fontenc}
\usepackage[latin9]{inputenc}
\usepackage{amstext}
\usepackage{esint}
\usepackage[normalem]{ulem}   

\newcommand{\beq}{\begin{equation}}
\newcommand{\eeq}{\end{equation}}

\begin{document}

\title{Beyond-Luttinger-Liquid thermodynamics of a one-dimensional contact repulsive Bose gas}

\author{Giulia De Rosi}
\email{giulia.derosi@icfo.eu}
\affiliation{Departament de F\'isica, Universitat Polit\`ecnica de Catalunya, Campus Nord B4-B5, 08034 Barcelona, Spain}
\affiliation{ICFO -- Institut de Ciencies Fotoniques, The Barcelona Institute
of Science and Technology, Av.~Carl Friedrich Gauss 3, 08860 Castelldefels (Barcelona), Spain}

\author{Pietro Massignan}
\email{pietro.massignan@upc.edu}
\affiliation{Departament de F\'isica, Universitat Polit\`ecnica de Catalunya, Campus Nord B4-B5, 08034 Barcelona, Spain}
\affiliation{ICFO -- Institut de Ciencies Fotoniques, The Barcelona Institute
of Science and Technology, Av.~Carl Friedrich Gauss 3, 08860 Castelldefels (Barcelona), Spain}

\author{Maciej Lewenstein}
\affiliation{ICFO -- Institut de Ciencies Fotoniques, The Barcelona Institute
of Science and Technology, Av.~Carl Friedrich Gauss 3, 08860 Castelldefels (Barcelona), Spain}
\affiliation{ICREA -- Instituci\'o Catalana de Recerca i Estudis Avan\c cats, Pg.~Llu\'is Companys 23, 08010 Barcelona, Spain}

\author{Grigori E. Astrakharchik}
\email{grigori.astrakharchik@upc.edu}
\affiliation{Departament de F\'isica, Universitat Polit\`ecnica de Catalunya, Campus Nord B4-B5, 08034 Barcelona, Spain}

\date{\today}

\begin{abstract}

We present a thorough study of the thermodynamics of a one-dimensional repulsive Bose gas, focusing in particular on corrections beyond the Luttinger-liquid description.
We compute the chemical potential, the pressure and the contact, as a function of temperature and gas parameter with exact thermal Bethe-Ansatz. In addition, we provide interpretations of the main features in the analytically tractable regimes, based on a variety of approaches (Bogoliubov, hard-core, Sommerfeld and virial).
The beyond Luttinger-liquid thermodynamic effects are found to be non-monotonic as a function of gas parameter.
Such behavior is explained in terms of non-linear dispersion and ``negative excluded volume'' effects, for weak and strong repulsion respectively, responsible for the opposite sign corrections in the thermal next-to-leading term of the thermodynamic quantities at low temperatures.
Our predictions can be applied to other systems including super Tonks-Girardeau gases, dipolar and Rydberg atoms, helium, quantum liquid droplets in bosonic mixtures and impurities in a quantum bath.

\end{abstract}

\maketitle

\section{Introduction}

Gapless systems in one spatial dimension often feature a linear phononic spectrum at low momenta, and this strongly constrains the low-temperature thermodynamics.
A unified description of the various quantum degeneracy regimes is then obtained within Luttinger Liquid (LL) theory, which relates the low-temperature properties of the system to the Luttinger parameter, i.e., the ratio of the Fermi velocity and the zero-temperature sound velocity, itself a function of the interaction strength \cite{Haldane1981, Voit1995,Giamarchi_book,Cazalilla2011}.
In the weakly-repulsive or Gross-Pitaevskii (GP) regime, a gas of bosons with short-range interactions admits a mean-field description \cite{Pitaevskii2016}.
In the opposite limit of very strong repulsion, the gas approaches the Tonks-Girardeau (TG) limit, where bosons become impenetrable and the system wave function can be mapped onto that of an ideal Fermi gas (IFG), resulting in indistinguishable thermodynamics \cite{Girardeau1960}.
Seminal experiments have explored this continuous interaction crossover in the past few years \cite{Paredes2004,Kinoshita2004,Tolra2004,Kinoshita2005,
Haller2011, Jacquim2011,Guarrera2012}.

The landscape of physical regimes in a one-dimensional (1D) Bose gas is even richer at higher temperature \cite{Petrov2000,
Vogler2013,
Salces-Carcoba2018}.
The correlation functions behave differently in the various regimes \cite{Kheruntsyan2003,Astrakharchik2003,Deuar2009,Panfil2014}, but those are hard to access experimentally.
Thermodynamic quantities can be measured more easily, but these generally exhibit a monotonic behavior to lowest order in temperature.
For example, the phononic excitations are responsible for the linear increase with temperature of the specific heat \cite{Pitaevskii2016}, and for the
quadratic growth of the chemical potential \cite{Lang2015,DeRosi2017}, for every interaction strength.
As the temperature is increased, however, higher momenta get explored and the deviation of the spectrum from the simple linear behavior becomes important \cite{Meinert2015, Fabbri2015}, resulting in a continuous structure bounded by two branches of elementary excitations \cite{Lieb21963}.
In the GP regime, the upper particlelike branch corresponds to the Bogoliubov spectrum \cite{Kulish1976} while the lower holelike one is instead associated with the dark soliton dispersion predicted by Gross-Pitaevskii theory \cite{Ishikawa1980}. In the opposite TG regime, the upper and the lower branches coincide with the particle and hole excitations of the ideal Fermi gas, respectively \cite{Pitaevskii2016}.
Such complex structure did not permit so far an easy physical interpretation of its effects on the corresponding thermodynamic behavior. This important gap is filled by the present work.
As we will demonstrate, the resulting thermal corrections are no longer monotonic and permit to classify the regimes of interaction.

In this paper, we provide a detailed study of the beyond Luttinger Liquid thermodynamics \cite{Imambekov2012, Fabbri2015} in a 1D Bose gas with short-range (contact) repulsion.
First, we solve numerically the thermal Bethe-Ansatz (TBA) equations within the Yang-Yang theory, which provide an exact answer to the problem at all temperatures $T$ and interaction strengths \cite{Yang1969,Yang1970}, and we compute key thermodynamic quantities, such as the chemical potential $\mu$, the pressure $P$, and the Tan's contact $\mathcal{C}$.
Then, we gain further insight into the problem by investigating analytically different tractable regions, including low- and high-temperatures, and weak- and strong-interactions.
We demonstrate that the Bogoliubov (BG) theory correctly describes thermodynamic properties at low temperatures and weak interactions.
For strong repulsion, we show that the leading interaction effects at both low and high temperatures stem from a ``negative excluded volume'' correction derived from the hard-core (HC) model.
Moreover, we demonstrate that the contact is proportional to the chemical potential in the GP limit at low temperatures and to the pressure in the TG regime for any $T$.
Finally, we show that the leading beyond-LL correction vs.~temperature in the investigated quantities (i.e. $\mu$, $P$, and $\mathcal{C}$) is negative in the GP regime and is positive in the TG limit. The same trend is also visible in the first correction to the leading classical gas contribution at high $T$.

\section{Model}
\label{Sec:Model}
The Hamiltonian of a 1D gas of $N$ bosons with contact repulsive interactions is given by
\begin{equation}
\label{Eq:H}
H = - \frac{\hbar^2}{2m}\sum_{i = 1}^N \frac{\partial^2}{\partial x_i^2} + g\sum_{i > j}^N\delta(x_i - x_j),
\end{equation}
where $m$ is the atom mass, $g = - 2 \hbar^2/(m a)$ is the coupling constant, and $a<0$ is the 1D $s$-wave scattering length.
The interaction strength is determined by the dimensionless quantity $\gamma = - 2/(n a)$ which depends on the gas parameter $n a$, with $n=N/L$ the linear density and $L$ the length of the system.
There is a crossover between the weak ($\gamma \ll 1$) and strong ($\gamma \gg 1$) interaction limits.
A peculiar feature of one dimension is that the high-density $n |a| \gg 1$ regime is described by the Bogoliubov theory contrarily to the usual three-dimensional case.
Instead, the low-density $n |a| \ll 1$ limit corresponds to a unitary Bose gas where the system \eqref{Eq:H} possesses the same thermodynamic properties of an IFG.

At zero temperature the system reduces to the Lieb-Liniger model, whose ground-state energy $E_0$, chemical potential $\mu_0 = (\partial E_0/\partial N)_{a, L}$ and sound velocity $v = \sqrt{n/m (\partial \mu_0/\partial n)_{a}}$ can be found from Bethe-Ansatz as a function of the interaction strength $\gamma$~\cite{Lieb1963, Lieb21963, Pitaevskii2016}.
The speed of sound smoothly changes from the mean-field value $v_{\rm GP} =\sqrt{g n/m}$ to the Fermi velocity $v_F = \hbar \pi n/m$ in the TG regime.

Within the canonical ensemble, the complete thermodynamics of the system is obtained starting from the Helmholtz free energy $A = E - TS$, with $E$ the energy and $S$ the entropy.
This allows for the calculation of the chemical potential
\begin{equation}
\label{Eq:mu}
\mu = (\partial A/\partial N)_{T,a,L},
\end{equation}
the pressure
\begin{equation}
\label{Eq:Pressure}
P=-(\partial A/\partial L)_{T,a,N} = n \mu - A/L,
\end{equation}
and the Tan's contact parameter \cite{Braaten2011,Yao2018}
\begin{equation}
\label{Eq:contact}
\mathcal{C}= (4 m/\hbar^2) \left(\partial A/\partial a  \right)_{T,L,N}.
\end{equation}
Simple considerations on scale invariance \cite{FetterBook, Barth2011} lead to a series of exact thermodynamic relations holding for any value of temperature and interaction strength (see Appendix \ref{Sec:thermodynamic relations}):
\begin{equation}
\label{Eq:thermodynamic identities}
- \frac{\mathcal{C}}{N} \frac{\hbar^2 a}{4 m} =
3\frac{A}{N} + 2 \frac{T S}{N} - \mu
= 2\frac{E}{N}-\frac{P}{n}.
\end{equation}

The chemical potential, the pressure, and the contact across the whole spectrum of temperature and interaction strength, as given by the solution of the thermal Bethe-Ansatz equations, are shown as symbols in Figs.~\ref{fig:mu}, \ref{fig:pressure}, and \ref{fig:contact}. 
The results are reported as ratios of the observables to their values given by the LL theory. With this choice, at low $T$, they all converge to unity for any value of the interaction strength $\gamma$, while, at higher $T$, any deviation from Luttinger Liquid line quantifies beyond-LL behavior which, instead, is strongly affected by $\gamma$. 
In these figures we rescale the temperature by the chemical potential at zero temperature $m v^2$, defining $\bar{T} \equiv k_B T/(m v^2)$. 
Since $m v^2$ is also the typical energy associated with phonons, depending on $\gamma$ through the sound velocity $v$, such temperature unit  is the proper one for the LL description holding in the whole interaction crossover.   
In the rest of the paper, we provide the understanding of dominant effects in the regimes which may be treated analytically.

\section{Chemical Potential}
\label{Sec:Chemical Potential}

\begin{figure}[t]
\begin{center}
\includegraphics[width=\columnwidth,angle=0,clip=]{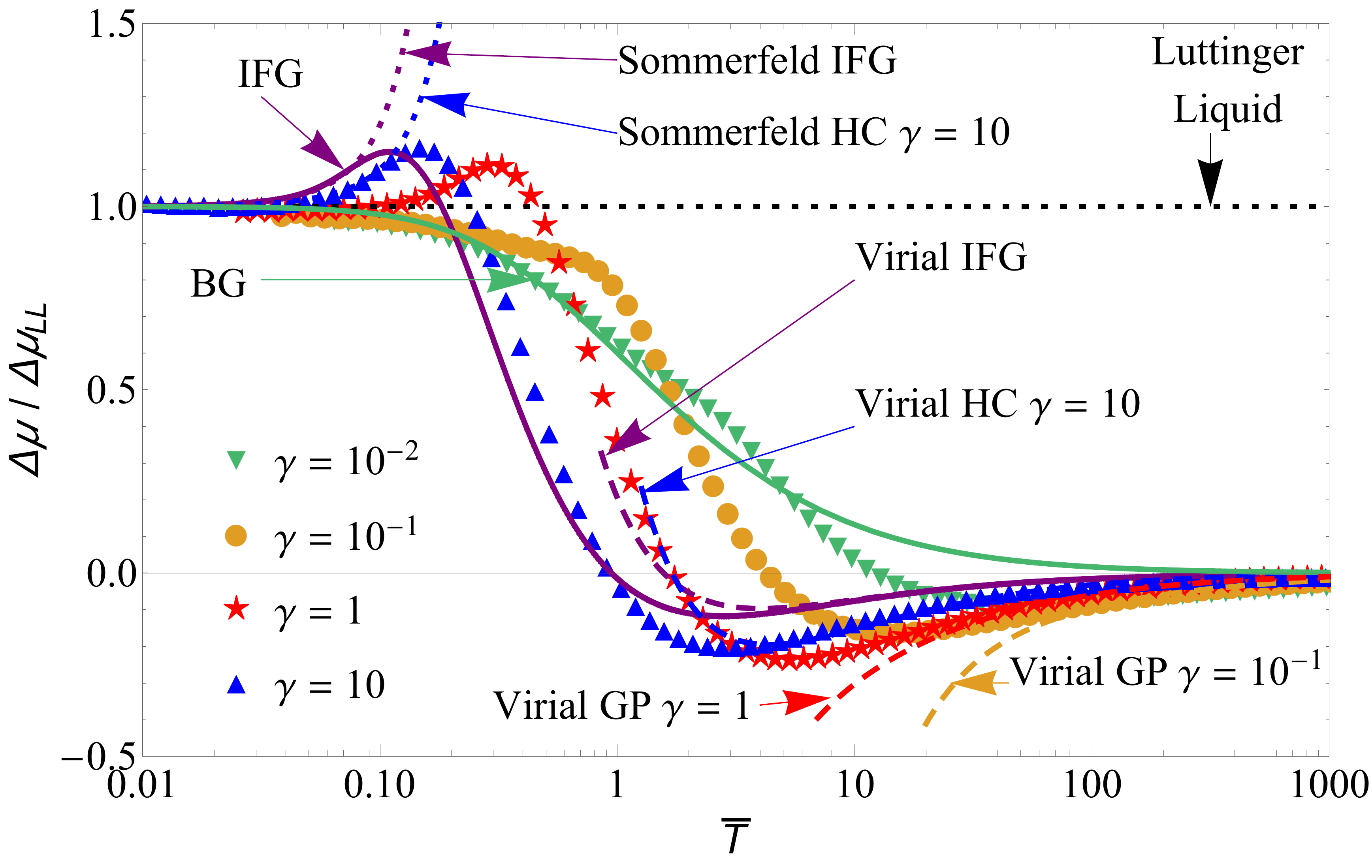}
\caption{Thermal shift of the chemical potential $\Delta\mu=\mu-\mu_0$ vs.~temperature. The symbols denote numerical TBA results for several interaction strengths $\gamma$, and the lines correspond to various theories described in the text.
The temperature is normalized to the typical energy associated with phonons, $\bar{T} = k_B T/(m v^2)$, and we measure shifts in terms of the LL result $\Delta\mu_{\rm LL}$, so that deviations from unity directly quantify beyond-LL effects. In these units the BG result, Eq.~\eqref{Eq:mu IBG}, is independent on $\gamma$.
}
		\label{fig:mu}
	\end{center}
\end{figure}

Let us start by considering weak interactions ($\gamma \ll 1$). At low temperatures $k_B T \ll mv^2$, the gas behaves like a quasicondensate, exhibiting features of superfluids \cite{Astrakharchik2004} with phononic excitations \cite{DeRosi2017}.
In this regime, the thermodynamics can be understood via Bogoliubov theory in terms of a gas of non-interacting bosonic quasi-particles \cite{Pitaevskii2016}. The thermal free energy is: 
\begin{equation}
\label{Eq:A IBG}
\Delta A_{\rm BG} = A - E_0 =  k_BTL \int_{-\infty}^{+\infty}\frac{dp}{2\pi \hbar}  \ln \left[1 - e^{-\beta \epsilon(p)} \right],
\end{equation}
where $\epsilon(p) = \sqrt{p^2 v^2 + \left[p^2/(2 m)\right]^2}$ is the $T = 0$ BG spectrum \cite{Lieb1963,Lieb21963}, which  depends on $\gamma$ through $v$.
From Eq.~\eqref{Eq:A IBG}, the temperature shift of the chemical potential is found to be
\begin{equation}
\label{Eq:mu IBG}
\Delta \mu_{\rm BG} = \mu - \mu_0 =  \left(\frac{\partial v}{\partial n } \right)_{L} \int_{-\infty}^{+\infty} \frac{dp}{2\pi\hbar} \frac{\partial \epsilon(p)}{\partial v} \frac{1}{e^{\beta \epsilon(p)}-1}.
\end{equation}
Within the LL theory, one retains only the phononic part of the BG dispersion, $\epsilon(p) \approx v |p|$, and obtains the universal result
$\Delta\mu_{\rm LL} = \bar{\mu}\bar{T}^2$, with $\bar{\mu} = \pi m^2 v^2 (\partial v/\partial n)_L /(6\hbar)$ \cite{DeRosi2017}. Expanding the BG spectrum to higher momenta, $\epsilon(p) \approx v|p| \left[ 1 + p^2/(8 m^2 v^2)\right]$, allows to compute the first correction beyond LL, which is $\mathcal{O}(\bar{T}^4)$ (see Appendix \ref{SubSec:BG low T}):
\begin{equation}
\label{Eq:mu BG small p}
\Delta \mu_{\rm BG} = \Delta\mu_{\rm LL} \left[ 1 - \pi^2 \bar{T}^2/4\right] + \mathcal{O}(\bar{T}^6).
\end{equation}
At $m v^2 \ll k_B T  \ll m v_F^2$, where $m v_F^2$ provides the degeneracy temperature, the gas is in the thermal degenerate state.

At even higher temperatures $k_B T \gg m v_F^2$, the gas behaves classically with negative chemical potential. In the GP regime, the dominant contribution to thermodynamics is determined by single-particle excitations. A reliable description in this case is provided by Hartree-Fock theory, which yields the chemical potential $\mu_{\rm GP} = \mu_{\rm IBG} + 2 g n$  \cite{Pitaevskii2016}, with $\mu_{\rm IBG}$ the chemical potential of the ideal Bose gas (IBG).
Hence, we perform the virial expansion of the equation of state in terms of a small effective fugacity $\tilde{z} = e^{\beta(\mu_{\rm GP} - 2 g n)} \ll 1$.
At leading order in temperature one obtains $\Delta \mu \approx \mu_{\rm GP}$, with:
\begin{equation}
\label{Eq:virial mu BG}
\mu_{\rm GP} = 
k_B T \left[\ln(n \lambda) - n \lambda/\sqrt{2}  \right]+\mathcal{O}(T^0)
\end{equation}
where $\lambda = \sqrt{2\pi\hbar^2/(m k_B T)}$ is the thermal wavelength.
Equation \eqref{Eq:virial mu BG}
is an expansion for small gas parameter $n \lambda \ll 1$ and it holds if $\lambda$ is much larger than the interaction range.
Equation~\eqref{Eq:virial mu BG} depends on the coupling constant $g$ only through the $\mathcal{O}(T^0)$ term (see Appendix \ref{SubSec:virial BG}).
For smaller $\gamma$ (i.e.~larger densities), higher values of $\bar{T}$ are needed for the agreement of Eq.~\eqref{Eq:virial mu BG} with TBA, as may be seen in Fig.~\ref{fig:mu}.

For strong interactions ($\gamma \gg 1$), the thermodynamics at any temperature may be addressed by making an analogy with the hard-core model \cite{Motta2016}. Its free energy is obtained from that of an ideal Fermi gas, subtracting from the system size an ``excluded volume'' $N a$, where $a$ is the diameter of the HC:
\begin{equation}
\label{Eq:A HS}
A_{\rm HC}(L) = A_{\rm IFG}(L \to \hat{L}\equiv L - N a).
\end{equation}
The scattering length $a$ is positive for hard-core potentials, and the available phase space is diminished by $Na$.
For the repulsive $\delta$-potential in Eq.~\eqref{Eq:H}, instead, the scattering length is negative and the phase space is increased by $N|a|$ effectively inducing ``negative excluded volume''.
Although the HC equation of state applies for $a>0$, its continuation to $a<0$ at $T=0$ differs from the Lieb-Liniger equation of state only by terms $\mathcal{O}(na)^4$, with such deviation attributed to the different phase shift dependence on the scattering momentum for $\delta$-function and hard-core potentials \cite{Astrakharchik2010}.
We find that the ``negative excluded volume'' correction turns out to be dominant for $\gamma\gg 1$ and permits to describe the thermodynamics of $\delta$-interacting gas even at high $T$, as shown in Fig.~\ref{fig:mu}.
Similarly, we expect that the ``positive excluded volume'' correction will be important for the thermodynamics of short-range gases with $a > 0$ in a strongly-correlated metastable state (super Tonks-Girardeau gas \cite{Astrakharchik2005,Haller2009}).

Following the Sommerfeld expansion of the IFG free energy  \cite{Ashcroft1976} (see Appendix \ref{SubSec:Sommerfeld IFG}), and taking into account the excluded volume correction for a HC gas through Eq.~\eqref{Eq:A HS}, we arrive to:
\begin{multline}
\label{Eq:mu HS Sommerfeld}
\mu_{\rm HC} = \hat{E}_F \left[\left(1 + \frac{2}{3} a \hat{n}  \right) + \frac{\pi^2\hat{\tau}^2}{12}  \left(1 + 2 a \hat{n}\right) \right.\\
\left.+ \frac{\pi^4\hat{\tau}^4}{36}  \left(1 + \frac{6}{5} a \hat{n} \right) + \frac{7\pi^6\hat{\tau}^6}{144} \left(1 + \frac{10}{9} a \hat{n}   \right) + \mathcal{O}(\hat{\tau}^8)\right] \ ,
\end{multline}
where $\hat{\tau} = k_B T/\hat{E}_F$, with an effective Fermi energy $\hat{E}_F = \hbar^2 \pi^2 \hat{n}^2/(2 m)$ depending on the rescaled density $\hat{n} = n/(1 - an)$ which takes into account the ``negative excluded volume'' and is applicable for $n |a| \ll 1$.
An alternative derivation of Eq.~\eqref{Eq:mu HS Sommerfeld} up to $\mathcal{O}(T^4)$-order was already presented in Ref.~\cite{Wadati2005}. 
We further note that our result of the IFG Sommerfeld expansion in Appendix \ref{SubSec:Sommerfeld IFG} corrects a minor misprint in the $\mathcal{O}(T^4)$-term of Ref.~\cite{Lang2015}.
From the $T = 0$ contribution of Eq.~\eqref{Eq:mu HS Sommerfeld}, we calculate the HC sound velocity $v_{\rm HC} = v_F/(1 - an)^2$.
By comparing Eqs.~\eqref{Eq:mu BG small p} and \eqref{Eq:mu HS Sommerfeld}, one notices that the $\mathcal{O}(T^2)$-phononic contribution is always positive, while quantum statistical effects are responsible for an opposite sign in BG and HC theories in the beyond-LL $\mathcal{O}(T^4)$-term.

At high $T$, we apply the virial expansion to the equation of state of an IFG, and we get the corresponding expansion of the free energy (see Appendix \ref{SubSec:virial IFG}). Using Eqs.~\eqref{Eq:A HS} and \eqref{Eq:mu}, we derive the virial expansion of the chemical potential of a hard-core gas:
\begin{equation}
\label{Eq:virial mu HS}
\mu_{\rm HC} = k_B T \left[\ln(\hat{n} \lambda) + a \hat{n} + \left(1 + \frac{a \hat{n}}{2}   \right) \frac{\hat{n} \lambda}{\sqrt{2}}  \right] + \mathcal{O}(T^0).
\end{equation}
Equations~\eqref{Eq:virial mu BG} and \eqref{Eq:virial mu HS} share the classical gas logarithmic term, while the second perturbative contribution $\mathcal{O}(n \lambda)$ exhibits an opposite sign emerging from quantum statistics, whose effects become important at lower $T$.
The TG regime ($\gamma = +\infty$) is recovered from the HC model when $a = 0$, and it possesses the same thermodynamic properties of an IFG.

\begin{figure}[t]
	\begin{center}
		\includegraphics[width=\columnwidth,angle=0,clip=]{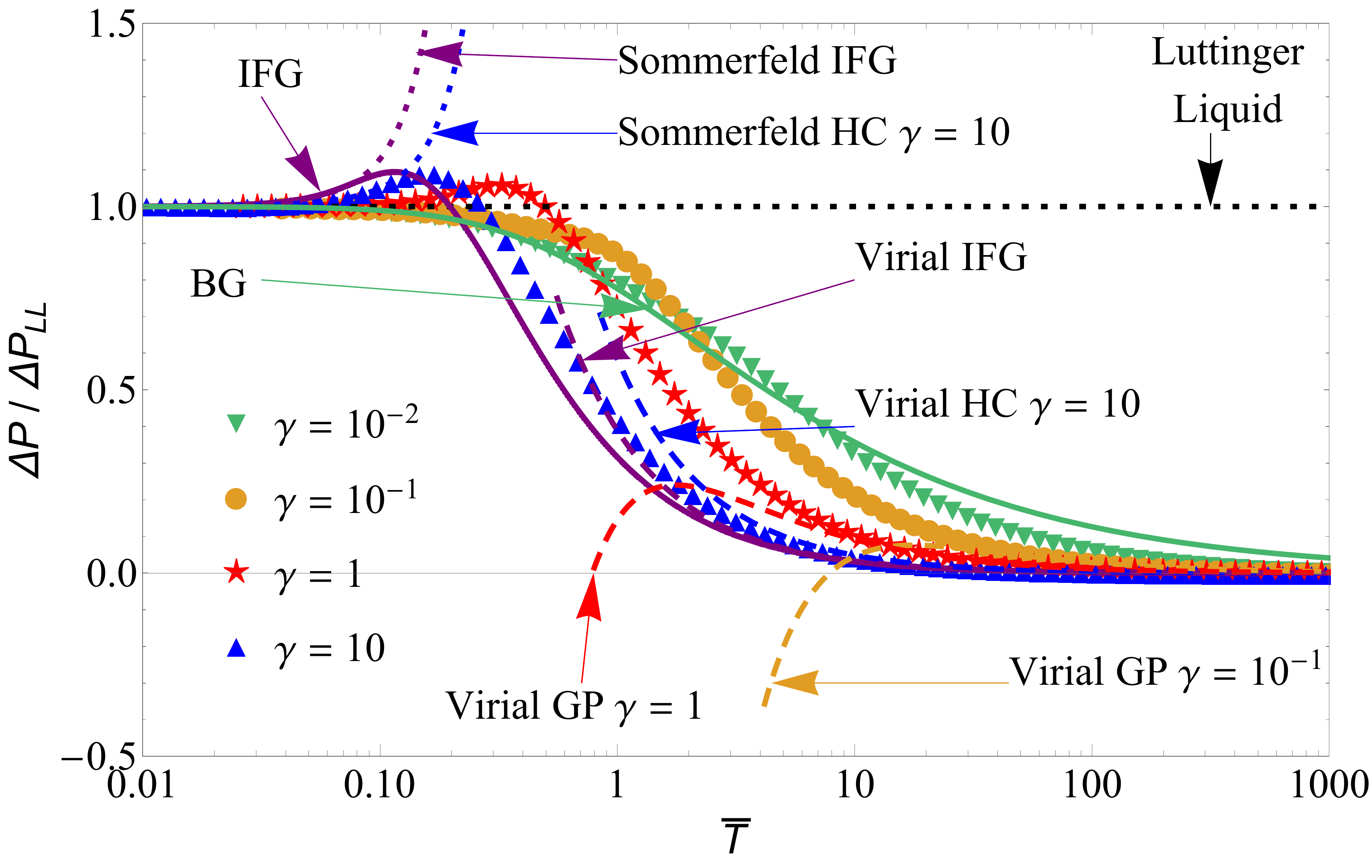}
		\caption{Thermal shift of the pressure. Symbols denote numerical TBA results, and lines correspond to various theories, as in Fig.~\ref{fig:mu}. Deviations from unity of $\Delta P/\Delta P_{\rm LL}$ indicate beyond-LL effects.
}
		\label{fig:pressure}
	\end{center}
\end{figure}

\section{Pressure}
\label{Sec:Pressure}

Let us now consider the thermal shift of the pressure, $\Delta P = P - P_0$, with $P_0 = n \mu_0 - E_0/L$ the pressure at $T = 0$.
Within BG theory, we start from Eqs.~\eqref{Eq:A IBG}-\eqref{Eq:mu IBG} and approximate $\epsilon(p) \approx v|p| \left[ 1 + p^2/(8 m^2 v^2)\right]$ to obtain (see Appendix \ref{SubSec:BG low T}):
\begin{equation}
\label{Eq:P BG small p}
\Delta P_{\rm BG} = \Delta P_{\rm LL} \left[1 - \frac{\pi^2\bar{T}^2}{20}  \frac{1 +
5 \chi_\gamma }{1 + \chi_\gamma } \right] +\mathcal{O}(\bar{T}^6).
\end{equation}
The leading-order (Luttinger liquid) result is $\Delta P_{\rm LL} =\bar{P}\bar{T}^2$, where
$\bar{P} = \pi m^2 v^3 \left(1 + \chi_\gamma\right)/(6\hbar)$ and $\chi_\gamma=\left[ \partial v/\partial n  \right]_L n/v$ depends on $\gamma$ through $v$.
The LL result in Eq.~\eqref{Eq:P BG small p} corrects a misprint in Ref.~\cite{Jiang2015}. The virial expansion is used to obtain the pressure in the GP regime at high $T$, resulting in:
\begin{equation}
\label{Eq:virial P BG}
P_{\rm GP} =  n k_B T [  1 - n \lambda/(2 \sqrt{2})] + \mathcal{O}(T^0) \ ,
\end{equation}
whose detailed derivation is reported in Appendix \ref{SubSec:virial BG}.

Within the HC approach, the Sommerfeld expansion of the free energy, Eq.~\eqref{Eq:A HS}, 
and Eqs.~\eqref{Eq:mu HS Sommerfeld} and \eqref{Eq:Pressure}, provide the following low-$T$ behavior of the pressure  
\begin{equation}
\label{Eq:P Sommerfeld HS}
P_{\rm HC} = \frac{2}{3}\hat{n} \hat{E}_F \left[1 + \frac{\pi^2}{4}\hat{\tau}^2 + \frac{\pi^4}{20}\hat{\tau}^4 + \frac{35\pi^6}{432}\hat{\tau}^6 + \mathcal{O}(\hat{\tau}^8)  \right],
\end{equation}
which is consistent with Ref.~\cite{Wadati2005} and includes the $\mathcal{O}(T^6)$-term.
The high-$T$ pressure is instead derived from the virial expansion of Eq.~\eqref{Eq:A HS}, and Eqs.~\eqref{Eq:virial mu HS} and \eqref{Eq:Pressure}: 
\begin{equation}
\label{Eq:P virial HS}
P_{\rm HC} =  \hat{n} k_B T  [1 + \hat{n} \lambda/(2\sqrt{2})] + \mathcal{O}(T^0).
\end{equation}
Equation~\eqref{Eq:P virial HS} provides the first quantum correction to the Tonks equation $P = \hat{n}k_B T$, which describes a classical HC gas \cite{Tonks1936,Mattis1993,Wadati2005}, and a higher-order interaction correction to the virial result of the  Yang-Yang theory \cite{Yang1970}, whose calculation is reported in Appendix \ref{Sec:YY virial pressure}.

\begin{figure}[t]
	\begin{center}
		\includegraphics[width=\columnwidth,angle=0,clip=]{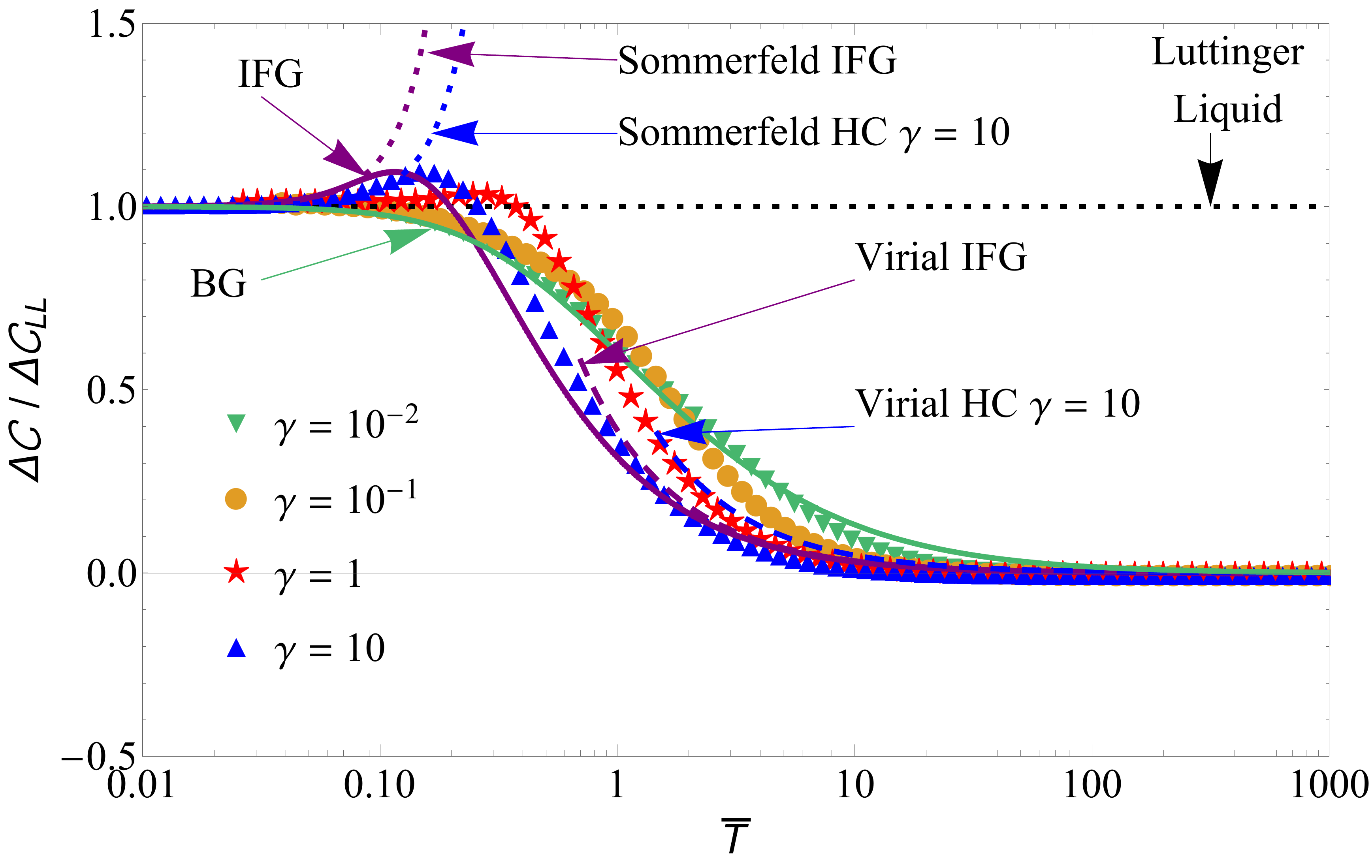}
		\caption{Thermal shift of the Tan's contact.~Symbols denote numerical TBA results, and lines correspond to various theories, as in Fig.~\ref{fig:mu}. Deviations from unity of $\Delta \mathcal{C}/\Delta \mathcal{C}_{\rm LL}$ indicate beyond-Luttinger-Liquid corrections.
		}
		\label{fig:contact}
	\end{center}
\end{figure}

\section{Tan's Contact}
\label{Sec:Tan's Contact}

In a system with zero-range interaction, the Tan's contact defined in Eq.~\eqref{Eq:contact} provides a relation between the equation of state and short-distance (large-momentum) properties, such as the interaction energy, the pair correlation function, 
and the relation between pressure and energy density \cite{Olshanii2003,Tan2008,Tan22008,Tan32008, Barth2011}, as shown for example in Eq.~\eqref{Eq:thermodynamic identities}.

Let us compute here the thermal contribution to the contact, $\Delta \mathcal{C}=\mathcal{C}-\mathcal{C}_0$,
where $\mathcal{C}_0=4 m/\hbar^2 (\partial E_0/\partial a)_{L, N}$ is the contact at $T=0$.
Within the BG theory, from Eq.~\eqref{Eq:A IBG} one obtains:
\begin{equation}
\label{Eq:contact IBG}
\Delta \mathcal{C}_{\rm BG} = (\bar{\mathcal{C}/}\bar{\mu}) \Delta\mu_{\rm BG},
\end{equation}
with $\bar{\mathcal{C}} = \pi m^3 v^2 N \gamma^2(\partial v/\partial \gamma)_n /(3 \hbar^3)$ entering in the LL result
$\Delta \mathcal{C}_{\rm LL}
= \bar{\mathcal{C}} \bar{T}^2$.
It can be shown that Eq.~\eqref{Eq:contact IBG} is consistent with Eq.~\eqref{Eq:thermodynamic identities}.
At high $T$ and in the GP regime, $\mathcal{C}$ does not depend on $T$, since within the Hartree-Fock approximation
the free-energy depends on $a$ only through the $T=0$ term (see Appendix \ref{SubSec:virial BG}).

With the HC approach, we find from Eqs.~\eqref{Eq:A HS} and \eqref{Eq:contact}:
\begin{equation}
\label{Eq:C HS}
\mathcal{C}_{\rm HC} = 4 m N P_{\rm HC}/\hbar^2.
\end{equation}
where the temperature dependence is encoded in $P_{\rm HC}$, which is given in Eqs.~\eqref{Eq:P Sommerfeld HS} and \eqref{Eq:P virial HS} for low- and high-T, respectively.
The relation between contact and pressure emerges from the HC excluded volume $\hat{L} = L - N a$, which transforms the $a$-dependence, Eq.~\eqref{Eq:contact}, in a $\hat{L}$-one: $\mathcal{C}_{\rm HC} \propto P_{\rm HC} = - (\partial A_{\rm HC} /\partial \hat{L})_{T, L, N}$, holding at any $T$. Eq.~\eqref{Eq:C HS} can be derived directly from Eq.~\eqref{Eq:thermodynamic identities} by using $E_{\rm HC} = \hat{L} P_{\rm HC}/2$.

\section{Experimental Considerations}
\label{Sec:Experimental Considerations}

The pressure, the chemical potential, the free energy, the energy and the entropy as a function of $T$ have been measured by using in-situ absorption imaging in three-dimensional ultracold gases \cite{Ho2009, Nascimbene2010b,Ku2012}.
A similar experimental technique has been applied to 1D Bose gas to extract the chemical potential as a function of temperature $T$ and interaction strength $\gamma$ \cite{Salces-Carcoba2018}, resulting in an excellent agreement with TBA.
Finally, Tan's contact parameter can be extracted from radio-frequency spectroscopy \cite{Wild2012, Sagi2012, Yan2019}, Bragg spectroscopy \cite{Hoinka2013} and from the large-momentum tail of the momentum distribution $n(k)$ \cite{Stewart2010, Chang2016}.

\section{Conclusions}
\label{Sec:Conclusions}

We provided a complete study of the chemical potential, the pressure and the contact as a function of temperature and interaction strength for a 1D Bose gas with repulsive contact interactions. Exact results were obtained within thermal Bethe-Ansatz theory and the main features were described analytically.
Beyond-Luttinger-Liquid effects were explained in terms of non-linear Bogoliubov dispersion relation for weak interactions and ``negative excluded volume'' for strong repulsion.
The beyond-LL effects are responsible for an opposite sign in the thermal next-to-leading term of the low-$T$ thermodynamic behavior, being negative in the GP limit and positive in the TG regime. The same trend is also visible in the first correction to the leading classical gas contribution at high $T$.
Finally, we found that the Tan's contact parameter is proportional to the chemical potential and to the pressure for weak and strong interactions, respectively.

In outlook, our work can stimulate further theoretical and experimental investigations aiming at the characterization of quantum degeneracy regimes, the beyond-Luttinger-Liquid physics and the microscopic nature of 1D Bose gases.
Our predictions are relevant for the investigation of the properties of impurities immersed in helium \cite{Bardeen1967}, in a 1D Bose gas \cite{Reichert2019} and in other 1D quantum liquids \cite{Recati2005, Schecter2014} as a function of $T$ and the interaction strength of the bath. 
Also, the knowledge of thermodynamics is crucial for the description of harmonically trapped gases \cite{Petrov2000,Vignolo2013,Xu2015,Yao2018}, especially for the investigation of breathing modes \cite{Moritz2003, Haller2009, Hu2014, Fang2014, Chen2015, Gudyma2015, DeRosi2016} whose frequency values are affected by the thermodynamic properties \cite{Astrakharchik2005bis, DeRosi2015}. 
Other interesting extensions of our work include multicomponent systems \cite{Yang2015,Decamp2016,Decamp22016} and configurations with a well-defined number of atoms \cite{Labuhn2016}.
The non-linear Bogoliubov dispersion effects are expected to be seen not only for contact interactions, but also in other short-range interacting systems provided that the density is high enough to be in the mean-field regime but not yet so large that the finite-range effects are visible. 
On the other hand, the ``excluded volume'' correction 
is expected to be applicable essentially to any short-range interacting system (integrable or not) at low density. For example, the ``excluded volume'' effects should be as well visible in $a>0 $ regime in
(i) metastable states of gas with short-range interactions, i.e.~for the super Tonks-Girardeau gas \cite{Astrakharchik2005,Haller2009}
(ii) gases with finite-range interactions such as dipolar atoms \cite{Arkhipov2005, Citro2007, Girardeau2012}, Rydberg atoms \cite{Osychenko2011}, bosonic $^4$He (liquid) in a certain density range \cite{Bertaina2016} and fermionic $^3$He (gas) at low densities \cite{Astrakharchik2014}.
Our results can be extended to 1D quantum liquid droplets in bosonic mixtures \cite{Petrov2015} in order to explore thermal effects. In particular,
1D enhances quantum fluctuations \cite{Petrov2016,Zin2018}, which are responsible for droplet stability, and it is achieved in current experiments \cite{Cheiney2018}.

\appendix

\section{Thermodynamic relations}
\label{Sec:thermodynamic relations}

In this Appendix, we provide details about the derivation of the thermodynamic relations.

Let us consider the following general expression of the free energy per particle required by dimensional analysis \cite{FetterBook, Barth2011}:
\begin{equation}
\label{Eq:A general}
\frac{A(T, a, n)}{N} \propto n^2 f\left(n a, \frac{T}{n^2} \right).\end{equation}
From Eq.~\eqref{Eq:A general}, one can deduce the scaling law
\begin{equation}
\label{Eq:A scaling law}
\frac{A(\ell^2T,\ell^{-1} a,\ell n)}{N} = \ell^2 \frac{A(T, a, n)}{N}
\end{equation}
where $\ell$ is an arbitrary, dimensionless parameter.
Taking the derivative of Eq.~\eqref{Eq:A scaling law} with respect to $\ell$ at $\ell = 1$ yields
\begin{multline}
\label{Eq:A Tan}
\Bigl[  2 T \left(\frac{\partial }{\partial T}\right)_{a, L, N} - a \left(\frac{\partial }{\partial a}\right)_{T, L, N} \\ + n \left(\frac{\partial}{\partial n}\right)_{T, a, L}   \Bigr] \frac{A(T, a, n)}{N} = 2 \frac{A(T, a, n)}{N} \ .
\end{multline}
From Eq.~\eqref{Eq:A Tan} and by using Eqs.~\eqref{Eq:mu} - \eqref{Eq:contact}, and $A = E - T S$ where $S = - (\partial A/\partial T)_{a, L, N}$, we find Eq.~\eqref{Eq:thermodynamic identities}.

\section{Weakly-interacting Bose gas}
\label{Sec:BG}

In this Appendix, we report the full calculation regarding the
low-and the high-temperature expansions of a weakly-interacting Bose gas. 

\subsection{Low-temperature expansion from non-linear Bogoliubov dispersion relation}
\label{SubSec:BG low T}
The low-momentum expansion of the Bogoliubov spectrum $x = \epsilon(p) = v|p| \left[ 1 + p^2/(8 m^2 v^2)\right]> 0$ may be inverted to find the only real and positive solution $p$. Hence, for the free energy,  Eq.~\eqref{Eq:A IBG}, we get the integral:
\begin{multline}
\label{Eq:integral A BG small p}
\int_{0}^{+\infty} dx
\frac{1}{v \left[1 + \frac{3 p^2(x)}{8\left(m v \right)^2}\right]} \ln\left(1 - e^{-\frac{x}{k_B T}} \right) \\ \approx - \frac{\pi^2}{6}\frac{\left(k_B T \right)}{v} + \frac{\pi^4}{120 } \frac{(k_B T)^3}{m^2 v^5},
\end{multline}
where the analytic solution may be found expanding the integrand for $|p| \ll m|v|$, which is justified at low temperatures. We find the low-$T$ expansion of the free energy, within the Bogoliubov theory:
\begin{multline}
\label{Eq:A BG small p}
\Delta A_{\rm BG} = A - E_0 \\ = - \frac{\pi}{6} \frac{\left(k_B T\right)^2 L}{\hbar v} \left[   1 - \frac{\pi^2}{20} \frac{\left(k_B T  \right)^2}{m^2 v^4} + \mathcal{O}(T^4)\right],
\end{multline}
where $E_0$ is the ground-state energy calculated within the Lieb-Liniger model at zero temperature \cite{Lieb1963}.

With a similar procedure for the chemical potential, Eq.~\eqref{Eq:mu IBG}, we get the integral:
\begin{multline}
\label{Eq:integral mu BG small p}
\int_{0}^{+\infty} dx  \frac{p^2(x)}{v x \left[1 + \frac{3 p^2(x)}{8(m v)^2}  \right]} \frac{1}{e^{\frac{x}{k_B T}} - 1} \\ \approx \frac{\pi^2}{6 } \frac{\left(k_B T\right)^2}{v^3} - \frac{\pi^4}{24} \frac{(k_B T)^4}{ m^2 v^7},
\end{multline}
which provides the low-$T$ expansion Eq.~\eqref{Eq:mu BG small p}.

\subsection{High-temperature virial expansion within the Hartree-Fock theory}
\label{SubSec:virial BG}

The equation of state for a 1D weakly interacting Bose gas with density $n$ and pressure $P$ can be derived from
 \begin{equation}
 \label{Eq:EOS virial BG}
 \begin{cases}
 n \lambda = g_{1/2}(\tilde{z}) \\
 P \lambda = g n^2 \lambda + k_B T g_{3/2}(\tilde{z}),
 \end{cases}
  \end{equation}
where $\tilde{z} = e^{\beta(\mu - 2 g n)}$ is the effective fugacity within the Hartree-Fock theory \cite{Pitaevskii2016}, the Bose functions are $g_\nu (z) = \sum _{i = 1}^{+\infty} z^i/i^\nu$, and the thermal wavelength is $\lambda = \sqrt{2\pi\hbar^2/(m k_B T)}$.

By inverting the expression for $n$ in Eq.~\eqref{Eq:EOS virial BG} in terms of $\tilde{z} \ll 1$, and by expanding it for small values of the gas parameter $n \lambda \ll 1$, we obtain
\begin{equation}
\label{Eq:virial z}
\tilde{z}= n \lambda - \frac{(n \lambda)^2}{\sqrt{2}}  + \frac{\sqrt{3} - 1}{\sqrt{3}} (n \lambda)^3 +\mathcal{O}[(n\lambda)^4].
\end{equation}
By using the definition of $\tilde{z}$ in Eq.~\eqref{Eq:virial z} and a further expansion for $n \lambda \ll 1$, we finally find the virial expansion of the chemical potential:
\begin{multline}
\label{Eq:virial mu BG full}
\mu_{\rm GP} = k_B T \Bigl[\ln\left(n \lambda \right) - \frac{n \lambda}{\sqrt{2}} + \frac{3\sqrt{3} - 4}{4\sqrt{3}} \left(n \lambda \right)^2  \\ - \frac{2\sqrt{3} - 5}{6\sqrt{2}} \left(n \lambda  \right)^3 +\mathcal{O}[(n\lambda)^4]\Bigr] + 2 g n.
\end{multline}
By considering Eq.~\eqref{Eq:virial z} in the equation of state of $P$, Eq.~\eqref{Eq:EOS virial BG}, we derive the expansion of the pressure:
\begin{multline}
\label{Eq:virial P BG full}
P_{\rm GP} = n k_B T \Bigl[ 1 - \frac{n \lambda}{2\sqrt{2}} \\ + \left( \frac{\gamma}{2 \pi} + \frac{3\sqrt{3} - 4}{6\sqrt{3}}  \right) \left( n \lambda \right)^2+\mathcal{O}[(n\lambda)^3]\Bigr].
\end{multline}
From Eqs.~\eqref{Eq:virial mu BG full}-\eqref{Eq:virial P BG full} and Eq.~\eqref{Eq:Pressure}, we calculate the high-$T$ behavior of the free energy:
\begin{multline}
\label{Eq:A virial BG}
A_{\rm GP} =  k_B T N \Bigl[\ln\left(n \lambda \right) - 1 - \frac{n \lambda}{2\sqrt{2}} \\ + \frac{3\sqrt{3} - 4}{4\sqrt{3}} \frac{\left(   n \lambda\right)^2}{3}
 +\mathcal{O}[(n\lambda)^3] \Bigr] + g n N.
\end{multline}
We notice that, at this level of approximation, the interactions appear in Eqs.~\eqref{Eq:virial mu BG full} and \eqref{Eq:A virial BG} only through their contribution at zero-temperature.

\section{Ideal Fermi gas}
\label{Sec:IFG}

In this Appendix, we provide the details about the Sommerfeld and the virial expansions of an ideal Fermi gas.

The equation of state of a 1D IFG with density $n$ and pressure $P$ can be derived from
\begin{equation}
 \label{Eq:EOS IFG}
 \begin{cases}
 n \lambda = f_{1/2}(z) \ , \\
 P \lambda = k_B T f_{3/2}(z),
\end{cases}
  \end{equation}
  where we have defined the fugacity $z = e^{\mu/(k_B T)}$ and the Fermi functions $  f_\nu(z) = \sum_{i = 1}^{+\infty} (-1)^{i-1} z^i/i^\nu$.

  \subsection{Low-temperature Sommerfeld expansion}
\label{SubSec:Sommerfeld IFG}

Let us briefly review the Sommerfeld expansion \cite{Ashcroft1976} which enables to calculate integrals of the form:
\begin{multline}
\label{Eq:Sommerfeld}
\int_{0}^{+\infty} d\epsilon H(\epsilon)f(\epsilon) = \int_{0}^\mu d\epsilon H(\epsilon) \\ + \sum_{i = 1}^{+\infty} a_i (k_B T)^{2i} \frac{d^{2i-1}}{d\epsilon^{2i-1}} H(\epsilon)|_{\epsilon = \mu},
\end{multline}
where
\begin{equation}
 f(\epsilon) = \frac{1}{e^{\frac{\epsilon - \mu}{k_B T}} + 1}
 \end{equation}
  is the Fermi-Dirac distribution.
Let us for example consider the 1D density of states of the IFG:
\begin{equation}
\label{Eq:Hepsilon}
 H(\epsilon) = \frac{1}{2\sqrt{E_F \epsilon}},
\end{equation}
where $E_F = k_B T_F = \hbar^2 \pi^2 n^2/(2 m)$ is the Fermi energy.
In Eq.~\eqref{Eq:Sommerfeld}, we have introduced the dimensionless number
\begin{equation}
 a_i = \left(2 - \frac{1}{2^{2(i - 1)}}  \right) \xi(2i)
\end{equation}
where $\xi(i)$ is the Riemann zeta function.

At very low $T$, the chemical potential of the IFG approaches the Fermi energy, hence we set $\mu \to E_F (1+ \delta)$ with $0\leq\delta \ll 1$. Then, we consider the Sommerfeld expansion, Eq.~\eqref{Eq:Sommerfeld}, up to the $\mathcal{O}(\delta^3)$-order, corresponding to the integer $i = 3$, and we require $\int_{0}^{+\infty} d\epsilon H(\epsilon) f(\epsilon) = 1$, ensuring the correct normalization. We solve the resulting equation for the real solution $\delta$ and we expand again in series, getting the chemical potential
\begin{equation}
\label{Eq:mu IFG Sommerfeld}
\mu_{\rm IFG} = E_F \left( 1 + \frac{\pi^2}{12} \tau^2 + \frac{\pi^4}{36} \tau^4 + \frac{7\pi^6}{144} \tau^6 + \mathcal{O}(\tau^8)   \right),
\end{equation}
with $\tau = k_B T/E_F$.

The Sommerfeld expansion, Eq.~\eqref{Eq:Sommerfeld},
allows one to obtain the low-temperature behavior $k_B T \ll \mu$ of the Fermi functions
$f_\nu\left(e^{\frac{\mu}{k_B T}}\right) \approx \frac{1}{\nu \Gamma(\nu)} \left(\frac{\mu}{k_B T}\right)^\nu \left[ 1 + \frac{\pi^2}{6}  \nu\left(\nu - 1 \right)\left( \frac{k_B T}{\mu} \right)^2   \right]$,
where $\Gamma(\nu)$ is the Euler Gamma function.
By using the latter expression in Eq.~\eqref{Eq:EOS IFG}, one recovers the result, Eq.~\eqref{Eq:mu IFG Sommerfeld}, and obtains the low-temperature expansion of the pressure:
\begin{equation}
\label{Eq:P IFG Sommerfeld}
P_{\rm IFG} = \frac{2}{3} n E_F \left[1 + \frac{\pi^2}{4} \tau^2 + \frac{\pi^4}{20} \tau^4 + \frac{35 \pi^6}{432}  \tau^6 + \mathcal{O}(\tau^8) \right].
\end{equation}
From Eqs.~\eqref{Eq:mu IFG Sommerfeld} and \eqref{Eq:P IFG Sommerfeld} and Eq.~\eqref{Eq:Pressure}, we calculate the low-$T$ expansion of the free energy:
\begin{equation}
\label{Eq:A Sommerfeld IFG}
A_{\rm IFG}= \frac{E_F N}{3} \left(1 - \frac{\pi^2}{4}\tau^2 - \frac{\pi^4}{60} \tau^4 - \frac{7 \pi^6}{432} \tau^6 + \mathcal{O}(\tau^8)   \right).
\end{equation}

\subsection{High-temperature virial expansion}
\label{SubSec:virial IFG}

By inverting the equation for the density $n$ for $z \ll 1$, Eq.~\eqref{Eq:EOS IFG}, and by expanding for $n \lambda \ll 1$, we find
\begin{equation}
\label{Eq:virial z IFG}
 z(n \lambda) = n \lambda + \frac{(n \lambda)^2 }{\sqrt{2}} +  \frac{\sqrt{3} - 1}{\sqrt{3}} (n\lambda)^3 + \mathcal{O}[(n \lambda)^4],
\end{equation}
from which, using the definition of $z$ and a further expansion for $n \lambda \ll 1$, we find the virial expansion of the chemical potential:
\begin{multline}
\label{Eq:virial mu IFG}
\mu_{\rm IFG}  = k_B T \Bigl[\ln(n \lambda) + \frac{n \lambda}{\sqrt{2}}  \\ + \frac{3\sqrt{3} - 4}{4\sqrt{3}}(n\lambda)^2 + \frac{2 \sqrt{3} - 5}{6\sqrt{2}}(n\lambda)^3 + \mathcal{O}[(n \lambda)^4] \Bigr].
\end{multline}
If we consider Eq.~\eqref{Eq:virial z IFG} in the equation of $P$, Eq.~\eqref{Eq:EOS IFG}, we derive the high-temperature behavior of the pressure
\begin{multline}
\label{Eq:virial P IFG}
P_{\rm IFG} = n k_B T \Bigl[ 1 + \frac{n \lambda}{2\sqrt{2}} \\ + \frac{3\sqrt{3} - 4}{6\sqrt{3}}\left( n \lambda \right)^2 + \frac{2\sqrt{3} - 5}{8\sqrt{2}} \left(n \lambda \right)^3  + \mathcal{O}[(n \lambda)^4] \Bigr],
\end{multline}
and of the free energy:
\begin{multline}
\label{Eq:virial A IFG}
A_{\rm IFG} =  k_B T N \Bigl[\ln\left(n \lambda\right) - 1 + \frac{n \lambda}{2 \sqrt{2}} \\ + \frac{3\sqrt{3} - 4}{4\sqrt{3}}\frac{(n \lambda)^2}{3} + \frac{2\sqrt{3} - 5}{6\sqrt{2}}\frac{(n \lambda)^3 }{4}  + \mathcal{O}[(n \lambda)^4]  \Bigr].
\end{multline}

\section{Yang-Yang virial expansion of the pressure}
\label{Sec:YY virial pressure}

In this Appendix, we report the calculation of the Yang-Yang virial expansion of the pressure in the weak and strong repulsion limits.

Let us consider the virial expansion of the pressure in the Yang-Yang model \cite{Yang1970}:
\begin{equation}
\label{Eq:P Yang-Yang}
\frac{P}{n k_B T} = 1 + \left[ \frac{1}{2\sqrt{2}} + e^{
x^2} \left(  \sqrt{\frac{2}{\pi}} \int_0^{x} dy \hspace{0.1cm} e^{-y^2} - \frac{1}{\sqrt{2}}    \right)    \right](n \lambda)
\end{equation}
where $x =  \gamma (n \lambda)/(2 \sqrt{2 \pi})$. Since in Eq.~\eqref{Eq:P Yang-Yang}, only $\mathcal{O}(n \lambda)$-terms are taken into account, we stop expansions at order $\mathcal{O}(x)$.

In the weakly interacting regime ($x \ll 1$), we get
\begin{equation}
\label{Eq:virial P BG Yang-Yang}
\frac{P_{\rm GP}}{n k_B T} = 1 - \frac{n \lambda}{2 \sqrt{2}} + \frac{\gamma }{2 \pi}(n \lambda)^2 + \mathcal{O}[(n \lambda)^3],
\end{equation}
which for $\gamma = 0$ reproduces only the first correction of the virial expansion of an ideal Bose gas.
We notice here that the expression we derived in Eq.~\eqref{Eq:virial P BG full} is more accurate than the one in Eq.~\eqref{Eq:virial P BG Yang-Yang}, as the former contains an extra $\mathcal{O}[(n \lambda)^2]$-term independent on $\gamma$, which only emerges from the next-to-leading order of the Yang-Yang virial expansion, Eq.~\eqref{Eq:P Yang-Yang}.

In the strong repulsive regime ($x \gg 1$), we obtain
\begin{equation}
\label{Eq:virial P Yang-Yang TG x}
\frac{P_{\rm HC}}{n k_B T} = \left(1 + na\right) +
\frac{n \lambda}{2 \sqrt{2}} + \mathcal{O}[(n \lambda)^2 ],
\end{equation}
which provides the first terms of the virial expansion of the hard-core model,
Eq.~\eqref{Eq:P virial HS}.

\begin{acknowledgments} 

G.~D.~R.~'s project has received funding from the European Union's Horizon 2020 research and innovation program under the Marie Sk\l odowska-Curie grant agreement UltraLiquid No 797684, under the Marie Sk\l odowska-Curie COFUND grant agreement PROBIST No 754510 and by the ``A. della Riccia'' Foundation.
G.~D.~R.~and M.~L.~acknowledge the Spanish Ministry MINECO (National Plan 15 Grant:
FISICATEAMO No.~FIS2016-79508-P, SEVERO OCHOA No.~SEV-2015-0522, FPI), European Social Fund, Fundaci\'o Cellex, Generalitat de Catalunya (AGAUR Grant No.~2017 SGR 1341 and CERCA/Program), ERC AdG OSYRIS and NOQIA, EU FETPRO QUIC, the National Science Centre and Poland-Symfonia Grant No.~2016/20/W/ST4/00314.
P.~M.~is supported by the ``Ram\'on y Cajal'' program.
G.~E.~A.~ and P.~M.~acknowledge funding from the Spanish MINECO (FIS2017-84114-C2-1-P). 
The Barcelona Supercomputing Center (The Spanish National Supercomputing Center - Centro Nacional de Supercomputaci\'on) is acknowledged for the provided computational facilities (RES-FI-2019-1-0006). 

\end{acknowledgments}

\bibliography{Bibliography}


 \renewcommand{\theequation}{S\arabic{equation}}
 \setcounter{equation}{0}
 \renewcommand{\thefigure}{S\arabic{figure}}
 \setcounter{figure}{0}
 \renewcommand{\thesection}{S\arabic{section}}
 \setcounter{section}{0}
 \onecolumngrid

\end{document}